\documentclass[12pt,english]{article}

\usepackage[T1]{fontenc}
\usepackage[latin9]{inputenc}
\usepackage{geometry}
\usepackage{color}
\usepackage{amstext}
\usepackage{graphicx}
\usepackage{setspace}
\usepackage{amssymb}
\usepackage{esint}
\usepackage{graphics,array,dcolumn}
\usepackage{xspace}
\makeatletter
\newcommand{\lyxaddress}[1]{
\par {\raggedright #1
\vspace{1.4em}
\noindent\par}
}

\makeatother

\usepackage{babel}

\begin{document}

\title{
\begin{flushright}
\small MZ-TH/11-04
\end{flushright}
\vspace{2cm}
Fluid Membranes and $2d$ Quantum Gravity}

\date{}


\author{Alessandro Codello%
\thanks{codello@sissa.it%
} $\;$and Omar Zanusso%
\thanks{zanusso@thep.physik.uni-mainz.de%
}}


\maketitle

\lyxaddress{\begin{center}
$^{*}$SISSA, Via Bonomea 265, 34136 Trieste, Italy
\par\end{center}}

\lyxaddress{\begin{center}
$^{\dagger}$Institut f\"ur Physik, Johannes Gutenberg-Universit\"at\\
Staudingerweg 7, D-55099 Mainz, Germany
\par\end{center}}

\begin{abstract}
We study the RG flow of two dimensional (fluid) membranes embedded
in Euclidean $D$-dimensional space using functional RG methods based
on the effective average action. By considering a truncation ansatz
for the effective average action with both extrinsic and intrinsic
curvature terms, we derive a system of beta functions for the running
surface tension $\mu_{k}$, bending rigidity $\kappa_{k}$, and Gaussian
rigidity $\bar{\kappa}_{k}$. We look for non-trivial fixed points
but we find no evidence for a crumpling transition at $T\neq0$. Finally,
we propose to identify the $D\rightarrow0$ limit of the theory with
two dimensional quantum gravity. In this limit, we derive new beta
functions for both cosmological and Newton's constants.
\end{abstract}

\section{Introduction}

The statistical mechanics of embedded two dimensional surfaces, or
membranes, is a complex and fascinating topic that spans different
areas of theoretical physics. In fact, embedded surfaces can be interpreted
both as the world-sheet of fundamental strings or, more realistically,
as describing chemically or biologically realized thermally fluctuating
two dimensional membranes. In the latter case, real world membranes
fall in two different classes: those with fixed connectivity are usually
called \emph{crystalline membranes}, while those which are not are
termed \emph{fluid membranes}. An effective action describing this
second class of membranes has been proposed a long time ago, \cite{Nelson_Piran_Weinberg_1988}
and is characterized by reparametrization invariance under coordinate
changes.

We describe a surface as a function $\mathbf{r}:\mathbb{R}^{2}\rightarrow\mathbb{R}^{D}$
which maps the point $x^{\alpha}$, $\alpha=1,2$, in a coordinate
patch to the embedding function $\mathbf{r}(x)$. At every point on
the surface, we define tangent vectors $\mathbf{r}_{\alpha}\equiv\partial_{\alpha}\mathbf{r}$
and $D-2$ normal unit vectors $\mathbf{n}^{i}$, $i=1,..,D-2$. Using
the induced metric tensor $g_{\alpha\beta}=\mathbf{r}_{\alpha}\cdot\mathbf{r}_{\beta}$
we can calculate the area of the surface $\int\sqrt{g}$ which is
also the simplest reparametrization invariant we can construct. To
every normal direction we associate an extrinsic curvature tensor
defined by $K_{\alpha\beta}^{i}=-\mathbf{r}_{\alpha}\cdot\mathbf{n}_{\beta}^{i}$.
Since the extrinsic curvature tensors are symmetric they can be diagonalized
at every point on the surface. This defines two principal curvatures
$k_{\pm}^{i}$ for every normal direction. The following two reparametrization
invariants are thus defined at each point (all surface indices are
raised using $g_{\alpha\beta}$): \begin{equation}
K^{i}K^{i}-K^{i\alpha\beta}K_{\alpha\beta}^{i}=2\sum_{i}k_{+}^{i}k_{-}^{i}\qquad\qquad K^{2}=K^{i}K^{i}=\sum_{i}(k_{+}^{i}+k_{-}^{i})^{2}\,,\label{A_1}\end{equation}
where we also defined the traces $K^{i}=g^{\alpha\beta}K_{\alpha\beta}^{i}$.
The invariants in (\ref{A_1}) are called, respectively, Gaussian
and mean curvature. The starting point of all modern differential
geometry was the observation made by Gauss, in his celebrated \emph{Theorema
Egregium}, that the Gaussian curvature is intrinsic, i.e. it does
not depend on how the surface is embedded. Gauss obtained the following
relation between the extrinsic curvature tensors and the (intrinsic)
Riemann curvature tensor: \begin{equation}
R_{\alpha\beta\gamma\delta}=K_{\alpha\gamma}^{i}K_{\beta\delta}^{i}-K_{\alpha\delta}^{i}K_{\beta\gamma}^{i}\,.\label{A_5}\end{equation}
 From (\ref{A_5}) we easily obtain the relation $R=K^{i}K^{i}-K^{i\alpha}K_{\alpha\beta}^{i}$,
showing that the Gaussian curvature is proportional to the (intrinsic)
Ricci scalar. Also, since we are working with two dimensional surfaces,
the Riemann tensor and the Ricci tensors are proportional to the Ricci
scalar.

Up to canonically irrelevant terms, we can consider the following
microscopic action describing fluid membranes \cite{Nelson_Piran_Weinberg_1988,Ambjorn_Durhuus_Jonsson_1997}:
\begin{equation}
S[\mathbf{r}]=\int d^{2}x\sqrt{g}\left\{ \mu+\frac{\kappa}{2}K^{2}+\frac{\bar{\kappa}}{2}R\right\} \,.\label{1}
\end{equation}
Here $\mu$ is the surface tension, which has dimensions of a squared
mass, $\kappa$ is the bending rigidity while $\bar{\kappa}$ is termed
Gaussian rigidity. Finally we recall that the integral over the whole
surface of the Ricci scalar is the Euler characteristic of the surface
$\chi=\frac{1}{4\pi}\int\sqrt{g}R$, thus we expect the coupling $\bar{\kappa}$
to be relevant only when we allow topology fluctuations. An action
of the sort of (\ref{1}) is also interesting from a string theory
point of view since the couplings $\kappa$ is a relevant extension
of the Nambu-Goto action, represented here by the first term in (\ref{1}),
and has been considered in \cite{Polyakov_1986}.

In this article we study the RG flow of the couplings appearing in
(\ref{1}) by functional RG group methods based on the effective average
action \cite{Berges_Tetradis_Wetterich_2002}. To preserve invariance
under reparametrizations, the effective average action is constructed
using the background field method. We employ heat kernel techniques
to extract the beta functions for the running couplings $\mu_{k},$
$\kappa_{k}$ and $\bar{\kappa}_{k}$, defined later in (\ref{F_1}),
for arbitrary embedding dimension $D$. For $D\ge3$ the RG flow represents
the phase diagram of fluid membranes at equilibrium and we can use
these beta functions to search for non-trivial fixed points of the
flow representing a possible crumpling transition at finite temperature,
where fluid membranes undergo a transition between a flat phase and
a totally crumpled one. This crumpling transition has been studied
within the effective average action approach in \cite{Kownacki_Mouhanna_2009,Essafi_Kownacki_Mouhanna_2010}.

Finally we advance the proposal to identify the $D\rightarrow0$ limit
of the theory with two dimensional quantum gravity, giving in this
way a novel and original construction of it. We show how to interpret
the idea that, for vanishing embedding dimension, only intrinsic fluctuations
of the membrane contribute effectively to the path integral. In particular,
we find that the flow of the extrinsic coupling vanish, showing that
extrinsic degrees of freedom freeze, while the intrinsic coupling
constants can be identified with the running cosmological and Newton's
constants. We then compare these new beta functions with those derived
employing the effective average action formalism directly to the purely
intrinsic theory of quantum gravity as first done in \cite{Reuter_1998}.

\section{Effective average action}

The effective average action is a scale dependent generalization of
the standard effective action that depends on an infrared scale $k$
and that interpolates smoothly between the microscopic action for
$k\rightarrow\infty$ and the effective action for $k\rightarrow0$.
When local symmetries are present, as here reparametrization invariance,
the effective average action is constructed employing the background
field method.

Under an infinitesimal reparametrization $x^{\alpha}\rightarrow x^{\alpha}+\epsilon^{\alpha}$
the embedding function $\mathbf{r}(x)$ transforms as:\begin{equation}
\mathbf{r}(x)\rightarrow\mathbf{r}(x)+\epsilon^{\alpha}\partial_{\alpha}\mathbf{r}(x)\,,\label{EAA_0}\end{equation}
while the microscopic action (\ref{1}) remains invariant. We parametrize
fluctuations around a background configuration $\bar{\mathbf{r}}$
in the following way:\begin{equation}
\mathbf{r}=\bar{\mathbf{r}}+\delta\mathbf{r}\,,\label{EAA_0.1}\end{equation}
on which we impose the background gauge-fixing condition $f_{\alpha}=0$,
with\begin{equation}
f_{\alpha}[\delta\mathbf{r};\bar{\mathbf{r}}]=\delta\mathbf{r}\cdot\partial_{\alpha}\bar{\mathbf{r}}\,.\label{EAA_0.2}\end{equation}
The gauge-fixing condition (\ref{EAA_0.2}) allows fluctuations only
in the normal directions. The ghost operator that follows from (\ref{EAA_0.2}),\begin{equation}
\frac{\delta f_{\alpha}[\delta\mathbf{r};\bar{\mathbf{r}}](x)}{\delta\epsilon^{\beta}(y)}=\delta(x-y)\left[\bar{g}_{\alpha\beta}+\partial_{\alpha}\delta\mathbf{r}\cdot\partial_{\beta}\bar{\mathbf{r}}\right]\,,\label{EAA_0.3}\end{equation}
is ultra-local and can thus be omitted in the following. We will enforce
the background gauge-fixing condition (\ref{EAA_0.2}) without exponentiation
of the delta functional $\delta[f_{\alpha}]$, i.e. we will integrate
only over normal fluctuations. We can thus set\begin{equation}
\delta\mathbf{r}=\nu^{i}\mathbf{n}^{i}\label{EAA_0.4}\end{equation}
by introducing $D-2$ fluctuation scalar fields $\nu^{i}$.

To construct the effective average action we choose the cutoff action
as follows:
\begin{equation}
\Delta S_{k}[\nu;\bar{\mathbf{r}}]=\frac{1}{2}\int d^{2}x\sqrt{\bar{g}}\nu^{i}R_{k}(\bar{\Delta})\nu^{i}\,.\label{EAA_1}
\end{equation}
Here $R_{k}(z)$ is the cutoff shape function and we defined $\Delta=-g^{\mu\nu}\nabla_{\mu}\nabla_{\nu}$.
Remember that the background metric $\bar{g}_{\alpha\beta}$ used
in (\ref{EAA_1}) is constructed employing the background field $\bar{\mathbf{r}}$, thus the action (\ref{EAA_1}) is quadratic in the fluctuation fields.
The effective average action can be defined implicitly by the following functional
integral:
\begin{equation}
e^{-\Gamma_{k}[\left\langle \nu\right\rangle ;\bar{\mathbf{r}}]}=\int D\chi\,\exp\left\{ -S_{k}[\chi+\left\langle \nu\right\rangle ;\bar{\mathbf{r}}]-\Delta S_{k}[\chi;\bar{\mathbf{r}}]+\int d^{2}x\sqrt{\bar{g}}\,\Gamma_{k}^{(1;0)}[\left\langle \nu\right\rangle ;\bar{\mathbf{r}}]\chi\right\} \,,\label{EAA_2}
\end{equation}
where $\left\langle \chi\right\rangle =0$. We can define the reparametrization
invariant effective average action $\bar{\Gamma}_{k}[\mathbf{r}]$
by setting $\left\langle \nu\right\rangle =0$ in the effective average
action defined in (\ref{EAA_2}):\begin{equation}
\bar{\Gamma}_{k}[\mathbf{r}]=\Gamma_{k}[0;\mathbf{r}]\,.\label{EAA_3}\end{equation}
We will study a truncation ansatz for (\ref{EAA_3}) in the next section.

The main virtue of the effective average action is that it satisfies
an exact RG flow equation. This can be derived by taking a scale derivative
of (\ref{EAA_2}) with respect to the {}``RG time'' $t=\log k/k_{0}$
and reads:\begin{equation}
\partial_{t}\Gamma_{k}[\left\langle \nu\right\rangle ;\bar{\mathbf{r}}]=\frac{1}{2}\textrm{Tr}\frac{\partial_{t}R_{k}(\bar{\Delta})}{\Gamma_{k}^{(2;0)}[\left\langle \nu\right\rangle ;\bar{\mathbf{r}}]+R_{k}(\bar{\Delta})}\,.\label{EAA_4}\end{equation}
From (\ref{EAA_4}) we immediately obtain the exact flow equation
for the reparametrization invariant effective average action defined
in (\ref{EAA_3}):\begin{equation}
\partial_{t}\bar{\Gamma}_{k}[\mathbf{r}]=\frac{1}{2}\textrm{Tr}\frac{\partial_{t}R_{k}(\Delta)}{\Gamma_{k}^{(2;0)}[0;\mathbf{r}]+R_{k}(\Delta)}\,.\label{EAA_5}\end{equation}
Note that the RG flow equation (\ref{EAA_5}) involves the Hessian
of the full functional $\Gamma_{k}[\left\langle \nu\right\rangle ;\bar{\mathbf{r}}]$
evaluated at $\left\langle \nu\right\rangle =0$. Thus the RG flow
equation (\ref{EAA_5}) is not a closed relation for $\bar{\Gamma}_{k}[\mathbf{r}]$.
In the next section we will make a truncation ansatz for $\bar{\Gamma}_{k}[\mathbf{r}]$
where we will have $\Gamma_{k}^{(2;0)}[0;\mathbf{r}]=\bar{\Gamma}_{k}^{(2)}[\mathbf{r}]$
and in this way we will be able to use (\ref{EAA_5}). Note also that
in principle a scale dependent wave function renormalization factor
$Z_{\nu,k}$ for the fluctuation fields $\nu^{i}$ should be present
in the cutoff action (\ref{EAA_1}). This generates in the denominator
of the rhs of the flow equations (\ref{EAA_4}) and (\ref{EAA_5})
a term of the form $-\eta_{\nu,k}R_{k}(\Delta)$, where $\eta_{\nu,k}=-\partial_{t}\log Z_{\nu,k}$
is the anomalous dimension of the fluctuation fields. This contribution
represents an RG improvement that reflects the fact the RG flow lives
in the enlarged theory space where the full functional $\Gamma_{k}[\left\langle \nu\right\rangle ;\bar{\mathbf{r}}]$
lives. For details on this point see \cite{Codello_PhD}. In this
paper we will drop this term, i.e we fix $\eta_{\nu,k}=0$.

\section{Truncation and beta functions}

In this paper we consider the following ansatz for the effective average
action, which is a scale dependent version of (\ref{1}):\begin{equation}
\bar{\Gamma}_{k}[\mathbf{r}]=\int d^{2}x\sqrt{g}\left\{ \mu_{k}+\frac{\kappa_{k}}{2}K^{2}+\frac{\bar{\kappa}_{k}}{2}R\right\} \,.\label{F_1}\end{equation}
Here $\mu_{k}$, $\kappa_{k}$ and $\bar{\kappa}_{k}$ are respectively
the running surface tension, the running bending rigidity and the
running Gaussian rigidity. The first thing to do to evaluate the rhs
of the flow equation (\ref{EAA_5}) is to compute the Hessian of the
action (\ref{F_1}). This can be done as usual by expanding the action
in powers of the fluctuation fields $\nu^{i}$ up to second order.
In particular we find the following result \cite{Forster_1986,Kleinert_1986,Capovilla_1994}:\begin{eqnarray}
\bar{\Gamma}_{k,ij}^{(2)}[\mathbf{r}] & = & \left(\kappa_{k}\Delta^{2}+\mu_{k}\Delta\right)\delta_{ij}+V_{ij}^{\alpha\beta}\nabla_{\alpha}\nabla_{\beta}+U_{ij}+O(K^{4})\nonumber \\
V_{ij}^{\alpha\beta} & = & \kappa_{k}\left[-\frac{1}{2}\left(K^{2}\delta_{ij}+2K_{i}K_{j}-4K_{i\gamma\delta}K_{j}^{\gamma\delta}\right)g^{\alpha\beta}+2K_{l}K_{l}^{\alpha\beta}\delta_{ij}\right]\nonumber \\
U_{ij} & = & \mu_{k}\left(K_{i}K_{j}-K_{i\alpha\beta}K_{j}^{\alpha\beta}\right)\,.\label{F_3}\end{eqnarray}
As we said in the previous section, within the truncation ansatz we
are considering we have that $\Gamma_{k}^{(2;0)}[0;\mathbf{r}]=\bar{\Gamma}_{k}^{(2)}[\mathbf{r}]$
and thus we can use the flow equation (\ref{EAA_5}). Note that the
running Gaussian rigidity $\bar{\kappa}_{k}$ does not enter in equation
(\ref{F_3}) since the relative operator, $\int\sqrt{g}R$, is proportional
to the Euler topological invariant of the surface and has thus vanishing
variations. Note also that the differential operator appearing in
(\ref{F_3}) is fourth-order in the covariant derivatives and the
fact that the running surface tension multiplies a second-order differential
operator. These structures stem from the fact that the metric and
the extrinsic curvature tensors are defined as second partial derivatives
of our degree of freedom $\mathbf{r}$. Note also that both $V_{ij}^{\alpha\beta}$
and $U_{ij}$ are proportional to $K^{2}$. After inserting (\ref{F_3})
in (\ref{EAA_5}) and expanding the denominator in powers of curvature
we obtain:\begin{eqnarray}
\partial_{t}\bar{\Gamma}_{k}[\mathbf{r}] & = & \frac{1}{2}\textrm{Tr}\frac{\partial_{t}R_{k}(\Delta)}{(\kappa_{k}\Delta^{2}+\mu_{k}\Delta)\delta_{ij}+V_{ij}^{\alpha\beta}\nabla_{\alpha}\nabla_{\beta}+U_{ij}+R_{k}(\Delta)}+O(K^{4})\nonumber \\
 & = & \frac{1}{2}\textrm{Tr}\left[G_{k}(\Delta)\partial_{t}R_{k}(\Delta)\right]-\frac{1}{2}\textrm{Tr}\left[G_{k}^{2}(\Delta)\partial_{t}R_{k}(\Delta)V_{ij}^{\alpha\beta}\nabla_{\alpha}\nabla_{\beta}\right]\nonumber \\
 &  & -\frac{1}{2}\textrm{Tr}\left[G_{k}^{2}(\Delta)\partial_{t}R_{k}(\Delta)U_{ij}\right]+O(K^{4})\,,\label{F_4.01}\end{eqnarray}
where we defined the following fourth-order regularized propagator:\begin{equation}
G_{k}(z)=\frac{1}{\kappa_{k}z^{2}+\mu_{k}z+R_{k}(z)}\,.\label{F_4.1}\end{equation}
The first trace in (\ref{F_4.01}) is easily calculated using the
standard heat kernel expansion (for more details see Appendix A of
\cite{Codello_Percacci_Rahmede_2009}):\begin{eqnarray}
\textrm{Tr}\left[G_{k}(\Delta)\partial_{t}R_{k}(\Delta)\right] & = & \frac{D-2}{4\pi}Q_{1}\left[G_{k}\partial_{t}R_{k}\right]\int d^{2}x\sqrt{g}\nonumber \\
 &  & +\frac{D-2}{24\pi}Q_{0}\left[G_{k}\partial_{t}R_{k}\right]\int d^{2}x\sqrt{g}R+O(K^{4})\label{F_6}\end{eqnarray}
The $Q$-functionals that appear in (\ref{F_6}) are defined in the
Appendix. To evaluate the second and the third trace in equation (\ref{F_4.01})
we use the algorithm proposed in \cite{Benedetti_Groh_Machado_Saueressig_2010,Anselmi_Benini_2007}.
We find the following results:\begin{equation}
\textrm{Tr}\left[G_{k}^{2}(\Delta)\partial_{t}R_{k}(\Delta)V_{ij}^{\alpha\beta}\nabla_{\alpha}\nabla_{\beta}\right]=-\frac{\kappa_{k}}{8\pi}Q_{2}\left[G_{k}^{2}\partial_{t}R_{k}\right]\int d^{2}x\sqrt{g}\left[DK^{2}-4R\right]+O(K^{4})\label{F_7}\end{equation}
\begin{equation}
\textrm{Tr}\left[G_{k}^{2}(\Delta)\partial_{t}R_{k}(\Delta)U_{ij}\right]=\frac{\mu_{k}}{4\pi}Q_{1}\left[G_{k}^{2}\partial_{t}R_{k}\right]\int d^{2}x\sqrt{g}R+O(K^{4})\,.\label{F_8}\end{equation}
In deriving (\ref{F_7}) and (\ref{F_8}) we used the following traces:\begin{equation}
U_{ii}=\mu_{k}R\qquad\qquad g_{\alpha\beta}V_{ii}^{\alpha\beta}=\kappa_{k}\left(DK^{2}-4R\right)\,.\label{F_5}\end{equation}
Inserting the ansatz that we made for the effective average action
(\ref{F_1}) in the lhs of the flow equation (\ref{EAA_5}) gives:\begin{equation}
\partial_{t}\bar{\Gamma}_{k}[\mathbf{r}]=\int d^{2}x\sqrt{g}\left\{ \partial_{t}\mu_{k}+\frac{1}{2}\partial_{t}\kappa_{k}K^{2}+\frac{1}{2}\partial_{t}\bar{\kappa}_{k}R\right\} \,.\label{F_11}\end{equation}
When we reinsert equations (\ref{F_6}), (\ref{F_7}) and (\ref{F_8})
in the rhs of (\ref{F_4.01}) and we compare with (\ref{F_11}), we
can easily extract the following beta functions\begin{eqnarray}
\partial_{t}\mu_{k} & = & \frac{D-2}{8\pi}Q_{1}\left[G_{k}\partial_{t}R_{k}\right]\nonumber \\
\partial_{t}\kappa_{k} & = & \frac{D}{8\pi}Q_{2}\left[G_{k}^{2}\partial_{t}R_{k}\right]\kappa_{k}\nonumber \\
\partial_{t}\bar{\kappa}_{k} & = & \frac{D-2}{24\pi}Q_{0}\left[G_{k}\partial_{t}R_{k}\right]-\frac{1}{4\pi}Q_{1}\left[G_{k}^{2}\partial_{t}R_{k}\right]\mu_{k}\nonumber \\
 &  & -\frac{1}{2\pi}Q_{2}\left[G_{k}^{2}\partial_{t}R_{k}\right]\kappa_{k}\,,\label{F_12}\end{eqnarray}
for the running surface tension, bending rigidity and Gaussian rigidity.
These beta functions, valid for any admissible cutoff shape function
$R_{k}(z)$, are the main result of this section.

\section{RG flow and the crumpling transition}

The system (\ref{F_12}) can be used to investigate the phase diagram
of termally fluctuating membranes and the eventual existence of a
crumpling transition. Once we choose a cutoff shape function $R_{k}(z)$ we can
evaluate explicitly the $Q$-functionals on the rhs of the beta functions
(\ref{F_12}). This can be done analytically if we employ $R_{k}(z)=(k^{4}-z^{2})\theta(k^{2}-z)$
as cutoff shape function. If we insert the integrals (\ref{C_4})
from the Appendix in equation (\ref{F_12}) we find the following
forms:\begin{eqnarray}
\partial_{t}\tilde{\mu}_{k} & = & -2\tilde{\mu}_{k}-\frac{D-2}{2\pi}\frac{1}{\sqrt{4-4\kappa_{k}+\tilde{\mu}_{k}^{2}}}\log\frac{2+\tilde{\mu}_{k}-\sqrt{4-4\kappa_{k}+\tilde{\mu}_{k}^{2}}}{2+\tilde{\mu}_{k}+\sqrt{4-4\kappa_{k}+\tilde{\mu}_{k}^{2}}}\nonumber \\
\partial_{t}\kappa_{k} & = & \frac{D}{2\pi}\left[\frac{\kappa_{k}\left(2-2\kappa_{k}-\tilde{\mu}_{k}\right)}{\left(\kappa_{k}+\tilde{\mu}_{k}\right)\left(4-4\kappa_{k}+\tilde{\mu}_{k}^{2}\right)}\right.\nonumber \\
 &  & \qquad\left.-\frac{\tilde{\mu}_{k}\kappa_{k}}{\left(4-4\kappa_{k}+\tilde{\mu}_{k}^{2}\right)^{3/2}}\log\frac{2+\tilde{\mu}_{k}-\sqrt{4-4\kappa_{k}+\tilde{\mu}_{k}^{2}}}{2+\tilde{\mu}_{k}+\sqrt{4-4\kappa_{k}+\tilde{\mu}_{k}^{2}}}\right]\nonumber \\
\partial_{t}\bar{\kappa}_{k} & = & \frac{D-8}{6\pi}+\frac{\tilde{\mu}_{k}(\tilde{\mu}_{k}+2)}{\pi\left(\kappa_{k}+\tilde{\mu}_{k}\right)\left(4-4\kappa_{k}+\tilde{\mu}_{k}^{2}\right)}\nonumber \\
 &  & \qquad+\frac{2\tilde{\mu}_{k}}{\pi\left(4-4\kappa_{k}+\tilde{\mu}_{k}^{2}\right)^{3/2}}\log\frac{2+\tilde{\mu}_{k}-\sqrt{4-4\kappa_{k}+\tilde{\mu}_{k}^{2}}}{2+\tilde{\mu}_{k}+\sqrt{4-4\kappa_{k}+\tilde{\mu}_{k}^{2}}}\,.\label{F_14}\end{eqnarray}
Note that we introduced in (\ref{F_14}) the dimensionless surface
tension $\tilde{\mu}_{k}=k^{-2}\mu_{k}$. The beta functions system
(\ref{F_14}) is the main result of this paper. Note that since $\bar{\kappa}_{k}$
is the coupling constant of a topological operator it does not appear
on the rhs of (\ref{F_14}) and its beta function is entirely determined
by the couplings $\kappa_{k}$, $\tilde{\mu}_{k}$. The system for
$\partial_{t}\tilde{\mu}_{k}$ and $\partial_{t}\kappa_{k}$ can thus
be solved independently. 

To look for crumpling transition, where the membrane undergoes a continuos
transition between a flat phase and a totally crumpled one, we need
to search for non-trivial fixed points of the beta function system
(\ref{F_14}). We can start by setting $\tilde{\mu}_{k}=0$ in (\ref{F_14})
obtaining the simple expressions:
\begin{equation}
\partial_{t}\kappa_{k}=\frac{D}{4\pi}\qquad\qquad\partial_{t}\bar{\kappa}_{k}=\frac{D-8}{6\pi}\,.\label{F_15}
\end{equation}
It can be shown, using the general properties that any admissible cutoff shape function must satisfy, that the coefficients in (\ref{F_15}) are scheme independent, i.e. they do not depend on the explicit form of $R_{k}(z)$. Since our truncation contains only those couplings already present
at the level of the microscopic action (\ref{1}) and since we set
the anomalous dimension of the fluctuation fields to zero, the beta
functions (\ref{F_15}) reproduce those derived using perturbation
theory \cite{Polyakov_1986,Forster_1986,Kleinert_1986}. In particular,
if we integrate the flow equations (\ref{F_14}) from the UV scale
$\Lambda$ to the IR scale $k$, we obtain:\begin{equation}
\kappa_{k}=\kappa_{\Lambda}-\frac{D}{4\pi}\log\frac{\Lambda}{k}\qquad\qquad\bar{\kappa}_{k}=\bar{\kappa}_{\Lambda}-\frac{D-8}{6\pi}\log\frac{\Lambda}{k}\,.\label{F_16}\end{equation}
These equations represents the one-loop renormalization of the bending
rigidity and of the Gaussian rigidity and show what we already know:
fluid membranes have a crumpling transition only at $T=0$. We arrive
at the same conclusion if we study the flow in the $(\tilde{\mu}_{k},\kappa_{k})$
plane. In particular we find no non-trivial fixed point for $\kappa_{k}$.
We conclude that there is no evidence for a crumpling transition at
$T\neq0$, at least within the truncation ansatz we are considering.

On the side of numerical simulations there are in the literature both results
in favor of a crumpling transition at finite temperature \cite{Baillie:1990zp} and in contrast  \cite{Bowick:1992hn}. One possible source for
this disagreement is the fact that different lattice versions of (\ref{1}) are employed by these authors \cite{Catterall:1988gw}. Our results are in agreement with the findings of  \cite{Bowick:1992hn}.

\section{Quantum gravity as the $D\rightarrow0$ limit}

Two dimensional quantum gravity can be formulated as a sum over equivalence classes of metrics on abstract two dimensional manifolds. The intrinsic metric represents the degree of freedom over which we integrate when performing the path integral and the action is constructed with intrinsic invariants built from it. Polyakov was the first to show \cite{Polyakov:1981rd,Morris:1989sb} that by minimally coupling $D$ scalar fields to two dimensional quantum gravity, we obtain a description of the dynamics of two dimensional membranes propagating (or fluctuating) in a $D$-dimensional Euclidean space.
%
%
In this setting it is easy to see that pure two dimensional quantum gravity should correspond to the theory of fluctuating membranes in the $D\rightarrow0$ limit.
Since in recent years the RG flow of quantum gravity in arbitrary dimensions has been extensively studied using the effective average action approach \cite{Reuter_1998,Codello_Percacci_Rahmede_2009}, we can try to compare these "intrinsic" beta functions 
with the analytical continuation to $D\rightarrow0$ of the beta functions we derived in the previous section.
In other words we explore the idea if it is possible to construct two dimensional quantum
gravity as the $D\rightarrow0$ limit of the statistical mechanics
of two dimensional membranes embedded in $\mathbb{R}^{D}$.
We stress that while in the intrinsic formulation the limit $D\rightarrow0$ just implies matter contributions to vanish, in the case of embedded membranes the limit $D\rightarrow0$ is purely formal and should be considered as an analytical continuation to be applied to universal quantities such as critical exponents.
%
%
%

To test this idea we will consider the beta function
system (\ref{F_14})  in the limit $D\rightarrow0$ together with the following identifications:
\begin{equation}
\mu_{k}=\frac{\Lambda_{k}}{8\pi G_{k}}k^{2}\qquad\qquad\bar{\kappa}_{k}=-\frac{1}{8\pi G_{k}}\,,\label{QG_0}
\end{equation}
where $\Lambda_{k}$ and $G_{k}$ are, respectively, the dimensionless
cosmological and Newton's constants. We then compare this new RG flow
for two dimensional quantum gravity with the one obtained in
\cite{Reuter_1998,Codello_Percacci_Rahmede_2009}. 

The first thing to note is that in (\ref{F_14}) the beta function
of the extrinsic coupling $\kappa_{k}$ is proportional to $D$ and
thus vanishes in the $D\rightarrow0$ limit. It follows that $\kappa_{k}=\kappa_{\Lambda}$
for any scale $k$ and thus $\kappa_{\Lambda}$ becomes a free parameter
of the theory. The beta functions for the two intrinsic couplings
$\tilde{\mu}_{k}$ and $\bar{\kappa}_{k}$ become, after the identification
(\ref{QG_0}) and in the limit $D\rightarrow0$, the following:
\begin{eqnarray}
\partial_{t}\Lambda_{k} & = & -2\Lambda_{k}-\frac{32}{3}G_{k}\Lambda_{k}+\frac{\Lambda_{k}^{2}\left(2+\frac{\Lambda_{k}}{8\pi G_{k}}\right)}{\pi\left(\kappa_{\Lambda}+\frac{\Lambda_{k}}{8\pi G_{k}}\right)\left[4-4\kappa_{\Lambda}+\left(\frac{\Lambda_{k}}{8\pi G_{k}}\right)^{2}\right]}\nonumber \\
 &  & +\frac{2\Lambda_{k}^{2}+8\pi G_{k}\sqrt{4-4\kappa_{\Lambda}+\left(\frac{\Lambda_{k}}{8\pi G_{k}}\right)^{2}}}{\pi\left[4-4\kappa_{\Lambda}+\left(\frac{\Lambda_{k}}{8\pi G_{k}}\right)^{2}\right]^{3/2}}\log\frac{2+\frac{\Lambda_{k}}{8\pi G_{k}}-\sqrt{4-4\kappa_{\Lambda}+\left(\frac{\Lambda_{k}}{8\pi G_{k}}\right)^{2}}}{2+\frac{\Lambda_{k}}{8\pi G_{k}}+\sqrt{4-4\kappa_{\Lambda}+\left(\frac{\Lambda_{k}}{8\pi G_{k}}\right)^{2}}}\nonumber \\
\partial_{t}G_{k} & = & -\frac{32}{3}G_{k}^{2}+\frac{G_{k}\Lambda_{k}\left(2+\frac{\Lambda_{k}}{8\pi G_{k}}\right)}{\pi\left(\kappa_{\Lambda}+\frac{\Lambda_{k}}{8\pi G_{k}}\right)\left[4-4\kappa_{\Lambda}+\left(\frac{\Lambda_{k}}{8\pi G_{k}}\right)^{2}\right]}\nonumber \\
 &  & -\frac{2G_{k}\Lambda_{k}}{\pi\left[4-4\kappa_{\Lambda}+\left(\frac{\Lambda_{k}}{8\pi G_{k}}\right)^{2}\right]^{3/2}}\log\frac{2+\frac{\Lambda_{k}}{8\pi G_{k}}-\sqrt{4-4\kappa_{\Lambda}+\left(\frac{\Lambda_{k}}{8\pi G_{k}}\right)^{2}}}{2+\frac{\Lambda_{k}}{8\pi G_{k}}+\sqrt{4-4\kappa_{\Lambda}+\left(\frac{\Lambda_{k}}{8\pi G_{k}}\right)^{2}}}\,.\label{QG_1}
\end{eqnarray}
The beta functions (\ref{QG_1}) for the dimensionless cosmological
and Newton's constants are a novel representation of the RG flow of
two dimensional quantum gravity.

We start to study the system (\ref{QG_1}) by considering the beta
function of Newton's constant at $\Lambda_{k}=0$, since in this limit the coefficient should become scheme
independent and thus comparable with the one calculated within the intrinsic formulation. For any cutoff shape function we find %
\begin{figure}
\begin{centering}
\includegraphics[scale=1.1]{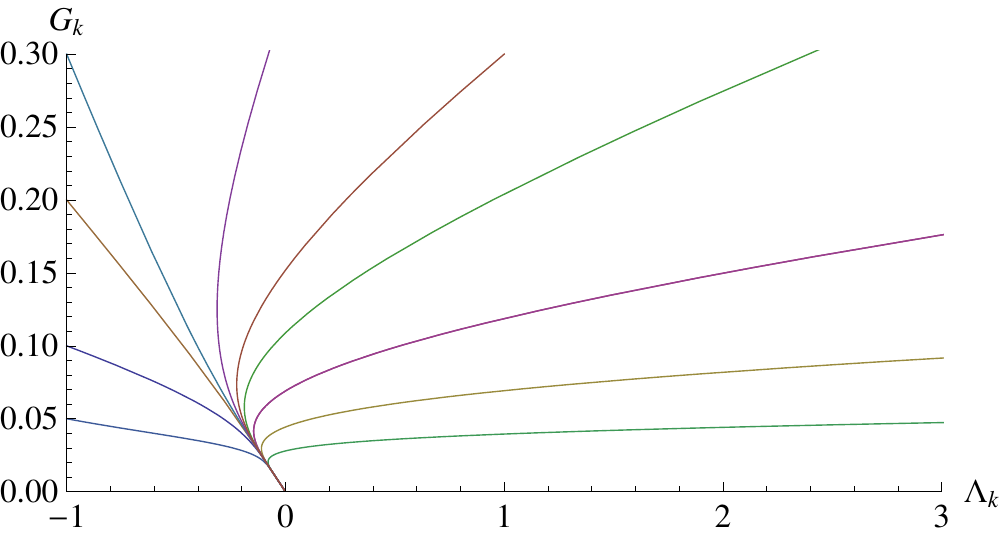}
\par\end{centering}

\caption{The RG flow in the $(\Lambda_{k},G_{k}$) plane obtained by integrating
the beta functions (\ref{QG_1}) for $\kappa_{\Lambda}=1$.}
\end{figure}
\begin{equation}
\partial_{t}G_{k}=-\frac{32}{3}G_{k}^{2}\,,\label{QG_3}\end{equation}
independently of the value that $\kappa_{\Lambda}$ assumes. The form
(\ref{QG_3}) can be compared with the beta function for Newton's
constant calculated using the intrinsic approach \cite{Reuter_1998,Codello_Percacci_Rahmede_2009}.
For $\Lambda_{k}=0$ and in the limit $d\rightarrow2$ (i.e. $\epsilon\rightarrow0$
if $d=2+\epsilon$) it is found:\begin{equation}
\partial_{t}G_{k}=-\frac{38}{3}G_{k}^{2}=\underbrace{-2}_{\textrm{graviton}}G_{k}^{2}\underbrace{-\frac{32}{3}}_{\textrm{ghost}}G_{k}^{2}\,.\label{QG_4}\end{equation}
In (\ref{QG_4}) we separated the contributions steaming from the
graviton and the ghost sectors. Our beta function (\ref{QG_3}) accounts
only for the ghost sector! In the calculation that we presented in
this paper we worked strictly in $d=2$ where the graviton sector,
based on the ansatz $\int\sqrt{g}R$, is not dynamical since it is
represented by a topological term. Thus it cannot give any contribution
to the beta function of Newton's constant. Differently, in the intrinsic
calculation \cite{Codello_Percacci_Rahmede_2009} one works in $d$-dimensions,
where the Hessian of $\int\sqrt{g}R$ is a well defined second-order
differential operator, and only at the level of the beta functions
one takes the limit $d\rightarrow2$, in this way obtaining (\ref{QG_4}).
Our result shows that the $d$-dimensional contribution to the beta
function of Newton's constant, when induced by a truncation containing
$\int\sqrt{g}R$, is not continuous in the limit $d\rightarrow2$.
In fact, for $d>2$ the inverse Hessian of $\int\sqrt{g}R$ has a
pole $(d-2)^{-1}$ which, in the effective average action calculation
\cite{Codello_Percacci_Rahmede_2009}, is cured by an appropriate
choice of the tensor structure of the cutoff action. Similar problems
have been encountered in perturbative calculations applied to $d=2+\epsilon$
quantum gravity \cite{Kawai_Ninomiya_1990,Jack_Jones_1991}.

For general values of $\Lambda_{k}$, one can ask at what value we
should fix the parameter $\kappa_{\Lambda}$. Note that, if we set
$\kappa_{\Lambda}=0$, the limit $\Lambda_{k}\rightarrow0$ of the
beta function of the dimensionless cosmological constant in (\ref{QG_1})
will diverge and so even the existence of the Gaussian fixed point
will be compromised. On the other side, to set $\kappa_{\Lambda}=0$
in our truncation ansatz (\ref{F_1}) means that we have to leave
out the kinetic term for extrinsic fluctuations represented by the
operator $\int\sqrt{g}K^{2}$. These facts show that it is necessary
to keep the extrinsic curvature invariant in the truncation ansatz
(\ref{F_1}) even if we are solely interested in the $D\rightarrow0$
limit, but any other non-zero value for $\kappa_{\Lambda}$ will be
a consistent choice. When we integrate the flow equations (\ref{QG_1})
for a given value $\kappa_{\Lambda}>0$ we obtain a phase portrait
similar to the one for $\kappa_{\Lambda}=1$ which is shown in Fig.
1. Obviously, we always find an UV attractive Gaussian fixed point,
showing that the theory is asymptotically free. Note that all the
RG flow trajectories that start with a positive $\Lambda_{k}$ approach
the Gaussian fixed point by passing trough values of negative dimensionless
cosmological constant. The same happens in the intrinsic case \cite{Codello_Percacci_Rahmede_2009}.
A nice property of the system (\ref{QG_1}) is that we are always
able to continue the flow towards the IR without encountering any
singularity, a problem that is instead present in the intrinsic counterpart
of (\ref{QG_1}). One can argue that if we were considering $d=2+\epsilon$
membranes, then the beta function for the (dimensionless) bending
rigidity in (\ref{F_14}), in the limit $D\rightarrow0$, will be
of the form $\partial_{t}\tilde{\kappa}_{k}=-\epsilon\tilde{\kappa}_{k}$
implying that the (dimensionless) bending rigidity flows towards infinity
in the IR. In view of this consideration we can consider the beta
functions (\ref{QG_1}) in the limit $\kappa_{\Lambda}\rightarrow\infty$.
We recover again (\ref{QG_3}) for the beta function of Newton's constant
and we find the following form for the beta function of the dimensionless
cosmological constant:\begin{equation}
\partial_{t}\Lambda_{k}=-2\Lambda_{k}-\frac{32}{3}\Lambda_{k}G_{k}\,.\label{QG_4.1}\end{equation}
The nice thing about the system composed of (\ref{QG_3}) and (\ref{QG_4.1})
is that flow trajectories with $\Lambda_{k}\ge0$ are present, feature
that we said is not present in the intrinsic case. 

The results of this section show that only the ghost contribution
to the gravitational RG flow is continuous in the variable $d$ while
the graviton contribution based on $\int\sqrt{g}R$ is not and is
strictly zero in $d=2$. We also understand that the role that in
the intrinsic calculations is played by the ghost sector, in the extrinsic
calculations is played by the extrinsic fluctuations, i.e. by those
invariants in the effective average action which are constructed using
the extrinsic curvature tensors $K_{\alpha\beta}^{i}$
\footnote{From a string theory point of view this shows that the Polyakov formulation,
which contains the ghost sector of two dimensional quantum gravity,
is equivalent to the Nambu-Goto formulation only if the extrinsic
curvature term $\int\sqrt{g}K^{2}$ is added to the action.}.
These results are very interesting and surprising if one considers
that the standard intrinsic formulation of two dimensional quantum
gravity and the extrinsic formulation that we proposed in this paper
differ in several respects: in the degrees of freedom one uses to
describe the system (i.e. $\mathbf{r}$ versus $g_{\mu\nu}$), in
the kind of gauge-fixing condition one employs, in the structure of
the ghost sector and in the order of the differential operators involved.
Still we find exact numerical agreement. This can be considered an
indication that the limit $D\rightarrow0$ ``exists''.

Finally, we speculate that if we were able to treat also the finite
terms arising on the rhs of the flow equation (\ref{F_4.01}), along
the lines of what was done in \cite{Codello_2010}, then in the $D\rightarrow0$
limit we should find that the effective average action will flow to
the non-local Polyakov action proper to the ghost system of two dimensional
quantum gravity. The other dynamical degree of freedom of two dimensional
quantum gravity is the conformal factor, but the truncation ansatz
(\ref{F_1}) gives no kinetic term to it. An obvious candidate for
this role is the non-local Polyakov action, but the treatment of a
truncation including this term is still difficult to handle within
the effective average action approach.

\section{Conclusions}

We have constructed the effective average action for two dimensional
fluid membranes embedded in a $D$-dimensional Euclidean space. We
used the background field method to preserve reparametrization invariance
of the theory and we derived the exact flow equation for this class
of theories.  We computed the beta functions for the running surface
tension, bending rigidity and Gaussian rigidity for any value of $D$.
For $D\ge3$ we found no evidence for a non-trivial fixed point, thus
confirming the expectation that fluid membranes are always in the
crumpled phase \cite{Nelson_Piran_Weinberg_1988,Bowick_Travesset_2001}.
It must be remembered that real world membranes are typically characterized by self-avoiding
interactions that penalize self-intersections of the membrane. It is thus important to consider
the effect of these non-local interactions on the phase diagram of the model \cite{Nelson_Piran_Weinberg_1988}.
The effective average action is probably one of the few analytical
formalisms where one can embark on the task of treating such interactions
analytically. This paper has to be seen as a first step in this direction.

Since two dimensional quantum gravity, when minimally coupled to $D$ scalar fields,
is equivalent to the theory of propagating membranes in $D$-dimensional Euclidean
space, we proposed to identify two dimensional
quantum gravity as the $D\rightarrow0$ limit of the statistical
mechanics of (fluid) membranes embedded in $\mathbb{R}^{D}$.
We tested this idea by considering this explicit limit in
the beta functions that we derived. We found first that the extrinsic
coupling (the bending rigidity) freezes, thus becoming a free parameter
in the beta functions of the dimensionless cosmological and Newton's
constants. More importantly, in the $\Lambda_{k}\rightarrow0$ limit,
the beta function of $G_{k}$ becomes scheme independent and we obtain
quantitative agreement with the corresponding beta function derived
in the intrinsic approach to two dimensional quantum gravity. More
precisely we found that we reproduced solely the ghost contributions.
The graviton contribution is strictly zero in $d=2$ and not continuous
in the $\epsilon\rightarrow0$ ($d=2+\epsilon$) limit since the operator
$\int\sqrt{g}R$ is a topological invariant in $d=2$. These results
are an indication of the equivalence between (intrinsic) two dimensional
quantum gravity and the statistical mechanics of fluctuating embedded
surfaces in the $D\rightarrow0$. The intuitive idea behind this is
that embedded membranes have both intrinsic and extrinsic fluctuations
and that these last can be {}``factored out'' by eliminating the
embedding space through the $D\rightarrow0$ limit. One expects to
be able to calculate universal properties of two dimensional quantum
gravity by this reformulation. It is tempting now to extend this conjecture
to higher dimensions: can we see $d$-dimensional quantum gravity
as the $D\rightarrow0$ limit of the statistical mechanics of $d$-dimensional
membranes embedded in a $D$-dimensional manifold (not necessarily
$\mathbb{R}^{D}$)? This will be particularly interesting in $d=4$
where intrinsic calculations point in the direction that quantum gravity
may be asymptotically safe \cite{Percacci_2009}, i.e. that the RG
group flow for this theory may have an UV attractive non-trivial fixed
point. Also, this reformulation can shade light on the role played
by the background metric in the effective average action formulation
of intrinsic $d$-dimensional quantum gravity \cite{Manrique_Reuter_2010}
since no background gauge-fixing and no background ghost actions are
present at the level of the bare action. All this will be the subject
of future work \cite{Codello_Zanusso_2011_bis}.

\section*{Acknowledgments}

We would like to thank R. Percacci for careful reading the manuscript
and for useful and stimulating discussions. The research of O.Z. is
supported by the Deutsche Forschungsgemein-schaft (DFG) within the
Emmy-Noether program (Grant No. SA/1975 1-1).

\appendix

\section{Threshold integrals}

To find the explicit expression for the beta functions (\ref{F_12})
we need to evaluate the $Q$-functionals. These are defined as follows:\begin{equation}
Q_{n}[f]=\frac{1}{\Gamma(n)}\int_{0}^{\infty}dz\, z^{n-1}f(z)\,,\label{C_1}\end{equation}
for $n>0$ and \begin{equation}
Q_{n}[f]=(-1)^{n}f^{(n)}(0)\,,\label{C_2}\end{equation}
for $n\le0$. We are interested in the cases where\begin{equation}
f(z)=G_{k}^{m}(z)\partial_{t}R_{k}(z)=\frac{\partial_{t}R_{k}(z)}{\left[az^{2}+bz+R_{k}(z)\right]^{m}}\,,\label{C_3}\end{equation}
for $m=1,2$. We will employ $R_{k}(z)=(k^{4}-z^{2})\theta(k^{2}-z)$
as cutoff shape function in order to perform all the integrals analytically.
Note that we have $\partial_{t}R_{k}(z)=4k^{4}\theta(k^{2}-z)$. We
find the following forms: \begin{eqnarray}
Q_{1}\left[G_{k}\partial_{t}R_{k}\right] & = & -\frac{4k^{2}}{\sqrt{4-4a+b^{2}}}\log\frac{2+b-\sqrt{4-4a+b^{2}}}{2+b+\sqrt{4-4a+b^{2}}}\nonumber \\
Q_{1}\left[G_{k}^{2}\partial_{t}R_{k}\right] & = & -4\frac{2+a\left(b-2\right)+\left(b-1\right)b}{(a+b)\left(4-4a+b^{2}\right)k^{2}}\nonumber \\
 &  & -8\frac{a-1}{\left(4-4a+b^{2}\right)^{3/2}k^{2}}\log\frac{2+b+\sqrt{4-4a+b^{2}}}{2+b-\sqrt{4-4a+b^{2}}}\nonumber \\
Q_{2}\left[G_{k}^{2}\partial_{t}R_{k}\right] & = & \frac{4\left(2-2a-b\right)}{\left(a+b\right)\left(4-4a+b^{2}\right)}\nonumber \\
 &  & +\frac{4b}{\left(4-4a+b^{2}\right)^{3/2}}\log\frac{2+b+\sqrt{4-4a+b^{2}}}{2+b-\sqrt{4-4a+b^{2}}}\nonumber \\
Q_{0}\left[G_{k}\partial_{t}R_{k}\right] & = & 4\,.\label{C_4}\end{eqnarray}
The integrals (\ref{C_4}) are used to obtain the beta functions given
in (\ref{F_14}).

\end{document}